\documentclass[10pt,conference]{IEEEtran}
\IEEEoverridecommandlockouts
\usepackage{cite}
\usepackage{amsmath,amssymb,amsfonts}
\usepackage{algorithmic}
\usepackage{graphicx}
\usepackage{textcomp}
\usepackage{xcolor}
\usepackage{booktabs}
\usepackage{subcaption}
\usepackage{ulem}
\usepackage{multirow}
\usepackage{listings}
\usepackage{color}
\definecolor{pblue}{rgb}{0.13,0.13,1}
\definecolor{pgreen}{rgb}{0,0.5,0}
\definecolor{pred}{rgb}{0.9,0,0}
\definecolor{pgrey}{rgb}{0.46,0.45,0.48}
\usepackage{wasysym}

\def\BibTeX{{\rm B\kern-.05em{\sc i\kern-.025em b}\kern-.08em
    T\kern-.1667em\lower.7ex\hbox{E}\kern-.125emX}}
\begin{document}

\newcommand{\appname}{{\sc {\sc EditSum}\ }}
\newcommand{\hx}[1]{\textcolor{blue}{#1}}

\title{{\sc EditSum}: A Retrieve-and-Edit Framework for Source Code Summarization}

\author{\IEEEauthorblockN{Jia Li \male} 
\IEEEauthorblockA{Key Lab of High Confidence Software \\
Technology, MoE (Peking University)\\
Beijing, China \\
lijia@stu.pku.edu.cn}
\and
\IEEEauthorblockN{Yongmin Li}
\IEEEauthorblockA{Key Lab of High Confidence Software \\
Technology, MoE (Peking University)\\
Beijing, China \\
liyongmin@pku.edu.cn}
\and
\IEEEauthorblockN{Ge Li* \thanks{* Corresponding authors}}
\IEEEauthorblockA{Key Lab of High Confidence Software \\
Technology, MoE (Peking University)\\
Beijing, China \\
lige@pku.edu.cn}
\and
\IEEEauthorblockN{Xing Hu}
\IEEEauthorblockA{School of Software Technology \\
Zhejiang University, Ningbo, China \\
xinghu@zju.edu.cn}
\and
\IEEEauthorblockN{Xin Xia}
\IEEEauthorblockA{Faculty of Information Technology \\
Monash University, Melbourne, Australia\\
Xin.Xia@monash.edu}
\and
\IEEEauthorblockN{Zhi Jin*}
\IEEEauthorblockA{Key Lab of High Confidence Software \\
Technology, MoE (Peking University)\\
Beijing, China \\
zhijin@pku.edu.cn}
}

\maketitle

\begin{abstract}
Existing studies show that code summaries help developers understand and maintain source code. Unfortunately, these summaries are often missing or outdated in software projects.
Code summarization aims to generate natural language descriptions automatically for source code.
According to Gros et al., code summaries are highly structured and have repetitive patterns (e.g. ``\textit{return true if...}'').
Besides the patternized words, a code summary also contains important keywords, which are the key to reflecting the functionality of the code.
However, the state-of-the-art approaches perform poorly on predicting the keywords, which leads to the generated summaries suffer a loss in informativeness.
To alleviate this problem, this paper proposes a novel retrieve-and-edit approach named {\sc EditSum} for code summarization. 
Specifically, {\sc EditSum} ﬁrst retrieves a similar code snippet from a pre-deﬁned corpus and treats its summary as a prototype summary to learn the pattern.
Then, {\sc EditSum} edits the prototype automatically to combine the pattern in the prototype with the semantic information of input code.
Our motivation is that the retrieved prototype provides a good start-point for post-generation because the summaries of similar code snippets often have the same pattern. The post-editing process further reuses the patternized words in prototype and generates keywords based on the semantic information of input code.
We conduct experiments on a large-scale Java corpus (2M) and experimental results demonstrate that {\sc EditSum} outperforms the state-of-the-art approaches by a substantial margin. The human evaluation also proves the summaries generated by {\sc EditSum} are more informative and useful. We also verify that {\sc EditSum} performs well on predicting the patternized words and keywords. 
\end{abstract}


\section{Introduction}
\label{sec:1}
During software development and maintenance, developers spend around 59\% of their time on program comprehension activities \cite{[2],[45],[46]}. A code summary provides a concise natural language description for a code snippet, which can help developers understand the program quickly and correctly \cite{[3]}.
Unfortunately, the code summaries are often mismatched, missing or outdated in the software projects~\cite{[47]}. Additionally, manually writing summaries during the development is time-consuming for developers. Therefore, it is important to explore automatic code summarization approaches.

Traditional approaches generate code summaries based on the template-based approaches and information retrieval (IR) based approaches. 
Template-based approaches~\cite{[3],[4]} ﬁrstly extract the keywords from the source code, and then fill the keywords into the predeﬁned templates to generate a code summary.
The IR-based approaches use code summaries of similar code snippets as outputs directly.
Among these IR-based approaches, they retrieve the similar code snippets by various similarity metrics~\cite{[6],[7]} from open-source software repositories in GitHub or software Q\&A sites~\cite{[8],[9]}.
Although the traditional approaches are simple, they have achieved good results. This is because code summaries are highly structured and contain many repetitive patterns, e.g., ``\textit{return true if...}'' and ``\textit{create a new...}''~\cite{[1]}.
The manually-crafted templates and retrieved summaries provide a lot of reusable patternized words, which play an key role in the code summaries.
However, for template-based approaches, manually defining templates is time-consuming and laborious, and requires a lot of expert experience. For IR-based approaches, there may be semantic inconsistencies between the retrieved summary and the input code.

With the development of deep learning, there is an emerging interest in applying neural networks for automatic code summarization.
Previous studies~\cite{[12],[13],[14]} often adopt the encoder-decoder architecture~\cite{[38]} to learn the mapping between words and even the grammatical structure from source code to natural language based on the large-scale corpus.
By virtue of the naturalness of the source code~\cite{[40],[41]}, these neural models can mine patterns for generating code summaries from a large corpus. 
Besides the patternized words, a code summary also contains important keywords, which have a low frequency in training data, but are the key to reflecting the functionality of source code (more details can be found in Section \ref{sec:2}).
However, the state-of-the-art nerual models~\cite{[12],[13],[14]} perform poorly on predicting keywords. For example, LeClair et al.~\cite{[14]} found 21\% summaries written by humans in the test set contain words with the frequency of less than 100, but only 7\% summaries generated by their proposed approach contain these words.
The lack of keywords leads to the generated summaries suffer a loss in informativeness, which have a negative impact on program comprehension.

Recently, Wei et al. \cite{[15]} and Zhang et al. \cite{[16]} proposed two retrieval-based neural models to address the problem of keywords. They used the IR techniques to get the similar code and its summary, and then input the retrieved results and the input code into the encoder. With the assistance of the retrieved summary, their models can accurately generate patternized words. However, their models only treated the retrieved results as auxiliary information and don't solve the problem of keywords.

In this paper, we propose a novel retrieve-and-edit approach {\sc EditSum} for code summarization.
The improvement by template-based approaches proves that the importance of the patterns in code summaries.
The improvement by IR-based approaches shows that the summaries of similar code snippets often have the same pattern.
So, we treat the summary of similar code as a prototype and extract the pattern from the prototype.
Considering the inconsistencies between the prototype and input code, we design a neural network  to further edit the prototype automatically based on the semantic information of input code.
Our motivation is that the pattern in a prototype tells the neural model ``how to say'' and the semantic information of input code tells the neural model ``what to say''.

{\sc EditSum} consists of two modules: a Retrieve module and an Edit module.
In the Retrieve module, given an input code snippet, we use IR techniques to retrieve the similar code snippet from a large parallel corpus and treat the summary of the similar code snippet as a prototype. 
Then, the Edit module generates a summary by fusing the pattern in prototype and semantic information of input code. 
Specifically, we propose a sequence-to-sequence (seq2seq) neural network to learn to revise the prototype based on the semantic differences of the input code and the similar code.
To represent the semantic differences, we calculate an edit vector by concatenating the weighted sums of insertion word embeddings (words in input code but not in similar code) and deletion word embeddings (words in similar code but not in input code).
After that, we revise the prototype summary conditioning on the edit vector to obtain a new summary.

To evaluate our approach, we conduct experiments on a real-world Java dataset. The dataset comes from the Sourcerer repository\footnote{https://www.ics.uci.edu/lopes/datasets/} and has been processed by LeClair et al. \cite{[14]}, including removing duplicates and dividing into training, validation, and test sets by projects. 
We employ the mainstream evaluation metric BLEU~\cite{[34]}, METEOR~\cite{[42]} and ROUGE~\cite{[43]} score that are widely used in summary generation task to evaluate the generated summaries.
Experimental results show that {\sc EditSum} performs substantially better than the IR-based baselines and outperforms the state-of-the-art neural baselines.
The human evaluation and qualitative analysis prove the summaries generated by {\sc EditSum} are informative and useful for developers to understand programs.
Besides, we verify that {\sc EditSum} not only accurately generates patternized words, but also generates more keywords.

Our main contributions are outlined as follows:
\begin{itemize}
    \item We propose a novel retrieve-and-edit approach, namely {\sc EditSum}, for code summarization. We use the summaries of similar code snippets as prototypes to assist in generating summaries.
    \item We design an effective editing module for summary generation, which can combine the pattern in prototype with the semantic information of code.
    \item We conduct extensive experiments to evaluate our approach on a large-scale Java dataset. The experimental results show that {\sc EditSum} substantially outperforms the state-of-the-art approaches.
\end{itemize}

\begin{table}[t]
\caption{The patterns of summaries in dataset.}
\label{tab:my-table}
\centering
\begin{tabular}{l|l}
\toprule
\multirow{5}{*}{Real Samples} & \textbf{write} \dashuline{a test finish} \textbf{to} \dashuline{the mesa logger}           \\
                              & \textbf{write} \dashuline{this tilemap} \textbf{to} \dashuline{an xml file}                \\
                              & \textbf{write} \dashuline{the buffer} \textbf{to} \dashuline{the output stream}            \\
                              & \textbf{write} \dashuline{grid data} \textbf{to} \dashuline{the geotiff file}              \\
                              & \textbf{write} \dashuline{cdl representation} \textbf{to} \dashuline{output stream}        \\
\hline
\multicolumn{1}{c|}{Pattern}   & write\_\_\_\_to\_\_\_\_                          \\
\hline
\multirow{5}{*}{Real Samples} & \textbf{this method sets} \dashuline{the help button visible}         \\
                              & \textbf{this method sets} \dashuline{the vaule of field}              \\
                              & \textbf{this method sets} \dashuline{a search argument for list} \\
                              & \textbf{this method sets} \dashuline{the client id}                   \\
                              & \textbf{this method sets} \dashuline{the range as a double}           \\
\hline
\multicolumn{1}{c|}{Pattern}   & this method sets\_\_\_\_\_\_\_\_        \\
\hline
\multirow{5}{*}{Real Samples} & \textbf{convert} \dashuline{an image} \textbf{to} \dashuline{an array of integer}         \\
                              & \textbf{convert} \dashuline{this ip packet} \textbf{to} \dashuline{a readable string}              \\
                              & \textbf{convert} \dashuline{a jingle description} \textbf{to} \dashuline{xml} \\
                              & \textbf{convert} \dashuline{the specified string} \textbf{to} \dashuline{a url}                   \\
                              & \textbf{convert} \dashuline{the date} \textbf{to} \dashuline{the given timezone}           \\
\hline
\multicolumn{1}{c|}{Pattern}   & convert\_\_\_\_to\_\_\_\_        \\
\bottomrule
\end{tabular}
\label{tab:motivation}
\end{table}

\textbf{Paper Organization.} The rest of this paper is organized as follows. Section \ref{sec:2} describes motivating examples. Section \ref{sec:3} presents our proposed approach. Section \ref{sec:4} and Section \ref{sec:5} describe the experimental setup and results. Section \ref{sec:6} and Section \ref{sec:7} discuss some results and describe the related work, respectively. Finally, Section \ref{sec:8} concludes the paper and points out future directions.

\section{Motivating Examples}
\label{sec:2}

A closer look at the code summarization dataset shows that patterns such as ``\textit{creates a new}'', ``\textit{returns true if}'', ``\textit{load into}'', ``\textit{convert into}'' are very frequent \cite{[1]}.
Table \ref{tab:motivation} shows some samples from the dataset provided by LeClair et al. \cite{[14]}.
The bold words are patternized words, and the dashed words denote the keywords.
Such a code summary can be regarded as composed of patternized words and keywords. 
The pattern ensures the readability of the summary, and the keywords reflect the functionality of the source code.
A good code summary should contains suitable patternized words and meaningful keywords.

\begin{figure}[t]
\centering
\includegraphics[width=\linewidth]{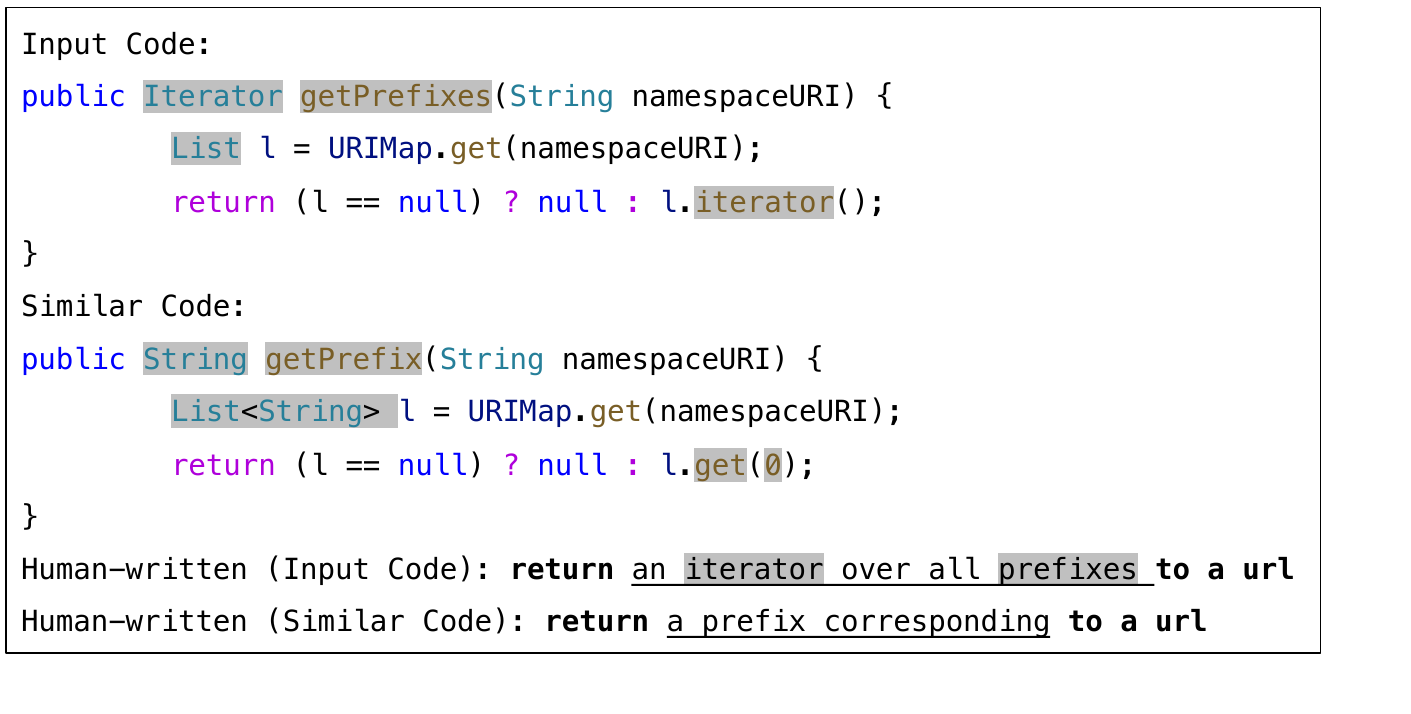}
\caption{An example of the input code and similar code.}
\label{fig:motivation}
\end{figure}

However, previous models perform well on predicting the patternized word, ignoring the importance of keywords.
As Figure~\ref{fig:motivation} shows, for the input code, we use the open-source search engine $Lucene$\footnote{https://lucene.apache.org/} to retrieve the most similar code snippet from the training corpus. The retrieval metric is based on the lexical level similarity of the source code. 

In Figure~\ref{fig:motivation}, the summaries of input code and similar code have the same pattern  (\textit{return...to a url}), but there are semantic differences between the similar code and input code. Although the two Java methods are lexically similar, the input code returns all prefixes, while the similar code returns a certain prefix. In Figure~\ref{fig:motivation}, the state-of-the-art neural model Rencos~\cite{[16]} can correctly predict the patternized words (e.g., return, to), but it performs poorly on keywords (e.g., prefixes). The code summaries generated by Rencos achieve high scores on the patternized words, but they do not clearly express the purposes of the programs. 

In this paper, we address that both pattern and keywords are important for a code summary. Inspired by previous studies, we
propose a retrieve-and-edit approach by combining the pattern in existing summaries and the semantic information of input code to generate informative summaries with suitable patterns.

\begin{figure*}[t]
\centering
\includegraphics[width=\linewidth]{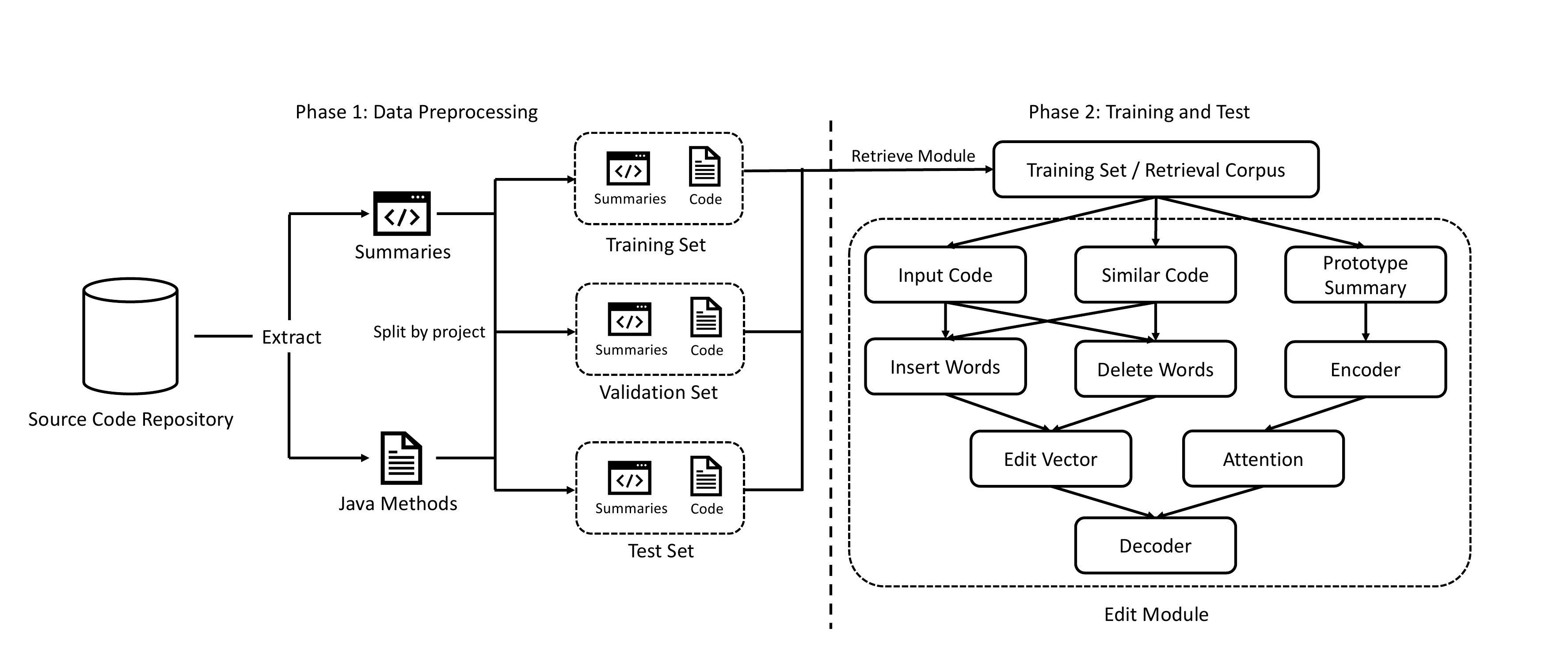}
\caption{The overall framework of our approach.}
\label{fig:framework}
\end{figure*}

\section{Proposed Approach}
\label{sec:3}
In this paper, we propose a retrieve-and-edit approach named {\sc EditSum} for source code summarization, which can combine the strengths of traditional approaches and neural models. 
The overall framework of our model is shown in Figure \ref{fig:framework}. 
Our approach {\sc EditSum} consists of a Retrieve module and an Edit module and generates a summary in three steps:

\textbf{Step 1:} Selecting a suitable prototype summary. We use a massive training set as the retrieval corpus. Given an input code, the Retrieve module uses the search engine to search for the similar code-summary pair from the corpus. The retrieval process is explained in Section \ref{sec:3.1}.

\textbf{Step 2:} Extracting the semantic information of the input code. In Figure \ref{fig:framework}, we mark the lexical differences between the two Java methods. We find that the different words between the two methods reflect their semantic differences to a certain extent, such as ``Iteration'' vs ``String'', and ``Prefixes'' vs ``Prefix''. Therefore, we calculate an edit vector based on the lexical differences between similar code and input code to represent their semantic differences. The details of this part is described in Section \ref{sec:3.2}.

\textbf{Step 3:} Combining the pattern in prototype with semantic information of input code. To this end, we design a neural edit module to revise the prototype based on the semantic differences between the input code and similar code. The details is presented in Section \ref{sec:3.2}.

\subsection{Retrieve Module}
\label{sec:3.1}

In our approach, the Retrieve module aims to retrieve the similar code-summary pair from a corpus given the input code.
Inspired by previous studies \cite{[15],[16]}, we choose the lexical-level similarity as retrieval metric. Specifically, we adopt $BM25$ \cite{[36]} as the similarity evaluation metric, which is a bag-of-words retrieval function to estimate the relevance of documents to a given query. Given a query and a document, based on TF-IDF~\cite{[39]}, the $BM25$ function calculates the term frequency in the document of each keyword in the query and multiplies it by the inverse document frequency of this term. The more relevant two documents have, the higher the value of $BM25$ score.
We leverage the open-source search engine $Lucene$ to build the Retrieve module. Since the size of the training set is quite large (over 1.9M), we use it as the retrieval corpus. We first tokenize the source code and summaries and process each code and summary pair into a document, add it to the index library, and store it on disk. 

As shown in Figure \ref{fig:framework}, we use different strategies to select prototypes for training and testing. 
In testing, we search for the most similar code from the training set and treat its summary as the prototype.
During the training phase, as we already know the targrt summary, we first retrieve top-20 code-summary pairs based on the summary similarity. Then, we reserve the retrieved summaries as prototypes whose $Jaccard$ similarity~\cite{[37]} to target summary in the range of [0.3, 0.7]. The $Jaccard$ similarity measures text similarity from a bag-of-words view, that is formulated as
\begin{equation}
J(A, B)=\frac{|A \cap B|}{|A \cup B|}
\end{equation}
where $A$ and $B$ are two bags of words and $|\cdot|$ denotes the number of elements in a collection.
The motivation behind ﬁltering out summaries with $Jaccard$ similarity $<$ 0.3 is the edit module performs well only if a prototype is lexically similar to its target summary \cite{[27]}. Besides, we hope the edit module does not copy the prototype so we discard summaries where the prototype and target summary are nearly identical (i.e. $Jaccard$ similarity $>$ 0.7). We do not use code similarity to construct training data, because similar code snippets may correspond to totally different summaries. 
This is not conducive to our model learning how to revise a prototype.
The preliminary experiments also show that constructing training data based on code similarity may cause the model to fail to converge.

\subsection{Edit Module}
\label{sec:3.2}

After that, the key step is to combine the pattern in the prototype and the semantic information of input code to generate a new summary. 
The structure of the Edit module is shown in Figure \ref{fig:edit_module}. Firstly, we utilize the prototype encoder to get the vector representation of prototype. Secondly, we compute the edit vector based on the lexical differences of two code snippets. The edit vector represents the semantic differences between the similar code and input code. Lastly, the summary decoder is used to generate a new summary conditioning on the prototype representation and edit vector.

\subsubsection{Prototype Encoder}
The prototype encoder takes the prototype $Y'$ as input. We first map the one-hot vector of a token $w'_i$ into a word embedding $y'_i$:
\begin{equation}
y'_{i}=W_{e}^{\top} w'_{i}, i \in [1,n]
\end{equation}
where $n$ is the length of prototype, $W_{e}$ is a trainable word embedding matrix. To leverage the contextual information, we use a bidirectional long short-term memory (Bi-LSTM) \cite{[33]} unit to process the sequence of word embeddings. At $i$-th time step, the hidden state $h_i$ of the Bi-LSTM can be represented by:
\begin{equation}
\overrightarrow{h}_{i}=\mathrm{LSTM}\left(\overrightarrow{h}_{i-1}, y'_{i}\right) ; \overleftarrow{h}_{i}=\mathrm{LSTM}\left(\overleftarrow{h}_{i+1}, y'_{i}\right)
\end{equation}
\begin{equation}
h_i = \left[\overrightarrow{h}_{i} \oplus \overleftarrow{h}_{i} \right]
\end{equation}
where $\oplus$ is a concatenation operation.
Finally, the prototype $Y'$ is transformed to vector representation $\{h_i\}_{i=1}^{n}$.

\subsubsection{Edit Vector}
The edit vector $z$ aims to reflect the semantic differences between the input code $X$ and similar code $X'$. Suppose that $X$ and $X'$ only differ by a single word $w$. Then one might think that the edit vector $z$ should be equal to the word embedding for $w$. Generalizing this intuition to multi-word edits, the multi-word insertions can be represented as the sum of the inserted word vectors, and similarly for multi-word deletions \cite{[27]}.

\begin{figure*}[t]
\centering
\includegraphics[width=0.9\linewidth]{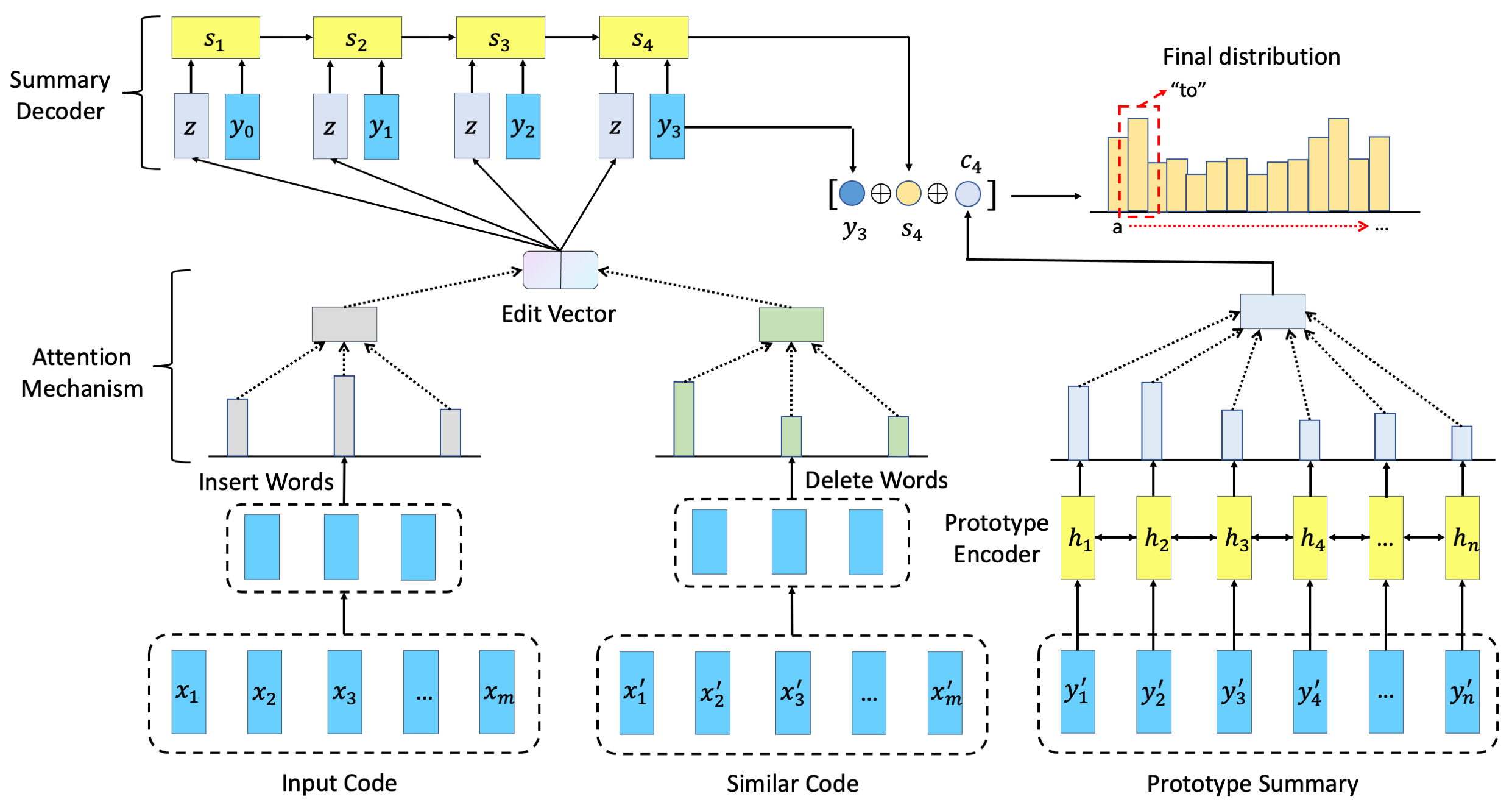}
\caption{The structure of the Edit module.}
\label{fig:edit_module}
\end{figure*}

As shown in Figure \ref{fig:edit_module}, we define $I=\{w \mid w \in X \wedge w \notin X'\}$ as a insertion word set, and $D=\{w' \mid w' \notin X \wedge w' \in X'\}$ as a deletion word set. Because different words inﬂuence the editing process unequally, we represent the differences between $X$ and $X'$ using the weighted sum of word embeddings:
\begin{equation}
f_{diff}\left(X, X'\right)=\sum_{w \in I} \alpha_w \Phi(w) \oplus \sum_{w' \in D} \beta_{w'} \Phi(w')
\end{equation}
where $\Phi(w)$ is the word embedding of word $w$ and $\oplus$ denotes a concatenation operation. $\alpha_w$ is the weight of a insertion word $w$, that is computed by the attention mechanism:
\begin{equation}
\alpha_{w}=\frac{\exp \left(e_{w}\right)}{\sum_{w \in I} \exp \left(e_{w}\right)}
\end{equation}
\begin{equation}
e_{w}=\mathbf{v}_{\alpha}^{\top} \tanh \left(\mathbf{W}_{\alpha}\left[\Psi(w) \oplus h_{n}\right]\right)
\end{equation}
where $\mathbf{v}_{\alpha}$ and $\mathbf{W}_{\alpha}$ are trainable parameters, and $h_n$ is the final hidden state of prototype encoder. $\beta_{w'}$ is obtained with a similar process.

Then we compute the edit vector $z$ by following linear projection, which can be regarded as a mapping from code differences to summary differences.
\begin{equation}
z=\tanh \left(\mathbf{W} \cdot f_{diff} +\mathbf{b}\right)
\end{equation}
where $\mathbf{W}$ and $\mathbf{b}$ are two trainable parameters.

\subsubsection{Summary Decoder}
After that, we revise the prototype based on the edit vector to get a new summary. The purpose of the summary decoder is to generate a new summary.

As shown in Figure \ref{fig:edit_module}, the decoder takes the prototype representation $\{h_i\}_{i=1}^{n}$ and the edit vector $z$ as inputs and generate a summary by a LSTM unit with attention. The hidden state of the decoder is compute by
\begin{equation}
s_{i}=\mathrm{LSTM}\left(s_{i-1}, y_{i-1} \oplus z\right)
\end{equation}
where $s_{i-1}$ means the previous hidden state of the decoder, $y_{i-1}$ is the ($i-1$)-th word embedding of ground-truth summary. We concatenate the edit vector to every input embedding of the decoder, so the edit information can be utilized in the entire generation process.

To introduce the information of the prototype, we then compute a context vector $c_i$ by attention mechanism, which is a weighted sum of prototype representation $\{h_i\}_{i=1}^{n}$:
\begin{equation}
c_{i}=\sum_{j=1}^{n} \eta_{i, j} h_{j}
\end{equation}
where attention weights are obtained by
\begin{equation}
\eta_{i, j}=\frac{\exp \left(e_{i, j}\right)}{\sum_{k=1}^{n} \exp \left(e_{i, k}\right)}
\end{equation}
\begin{equation}
e_{i, j}=\mathbf{v}_{\eta}^{\top} \tanh \left(\mathbf{W}_{\eta}\left[h_{j} \oplus s_{i}\right]\right)
\end{equation}
where $\mathbf{v}_{\eta}$ and $\mathbf{W}_{\eta}$ are two trainable parameters. Based on the previous word $y_{i-1}$, hidden state of the decoder $s_i$ and the context vector $c_i$ from prototype, our model compute the probability of $i$-th token $y_i$:
\begin{equation}
\label{equ:p}
p\left(y_{i}\right)=\operatorname{softmax}\left(\mathbf{W}_{\mathbf{p}}\left[y_{i-1} \oplus s_{i} \oplus c_{i}\right]+\mathbf{b}_{\mathbf{p}}\right)
\end{equation}
where $\mathbf{W}_{\mathbf{p}}$ and $\mathbf{b}_{\mathbf{p}}$ are two trainable parameters.

\subsection{Loss Function}
During training, {\sc EditSum} takes a token sequence of the input code $X$, a summary of the input code $Y$, a token sequence of the similar code $X'$, and the prototype $Y'$ as inputs, respectively. We optimize parameters of {\sc EditSum} by maximizing the probability of $p(Y|z, Y')$. The overall objective function of our model is to minimize the following loss function:
\begin{equation}
\mathcal{L}(\theta)=-\sum_{i=1}^{N} \sum_{t=1}^{L} \log P\left(y_{t}^{i} \mid z_i,Y'_i,y_{<t}^i\right)
\end{equation}
where $\theta$ is all trainable parameters. $N$ is the total number of training instances and $L$ is the length of each ground-truth summary. 

During testing, we utilize the prototype encoder to represent prototypes and compute edit vectors. Then, the summary decoder is used to generate directly a summary conditioning on the prototype representation and edit vector in Equation (\ref{equ:p}).

\begin{table}[t]
\small
\centering
\caption{The statistics of datasets.}
\begin{tabular}{lrrr}
\toprule \
\textbf{Dataset} & \textbf{Train} & \textbf{Valid} & \textbf{Test} \\ \midrule
Count & 1,954,807 & 104,273 & 90,908 \\
Avg. tokens in code & 29.67 & 29.68 & 30.17\\
Avg. tokens in summary & 7.594 & 7.710 & 7.654\\
\bottomrule 
\end{tabular}\vspace{-.4cm}
\label{tab:dataset}
\end{table}

\begin{figure}[t]
\centering
\begin{minipage}{\linewidth}
    \begin{subfigure}[t]{0.5\linewidth}
    \centering
    \includegraphics[width=\textwidth]{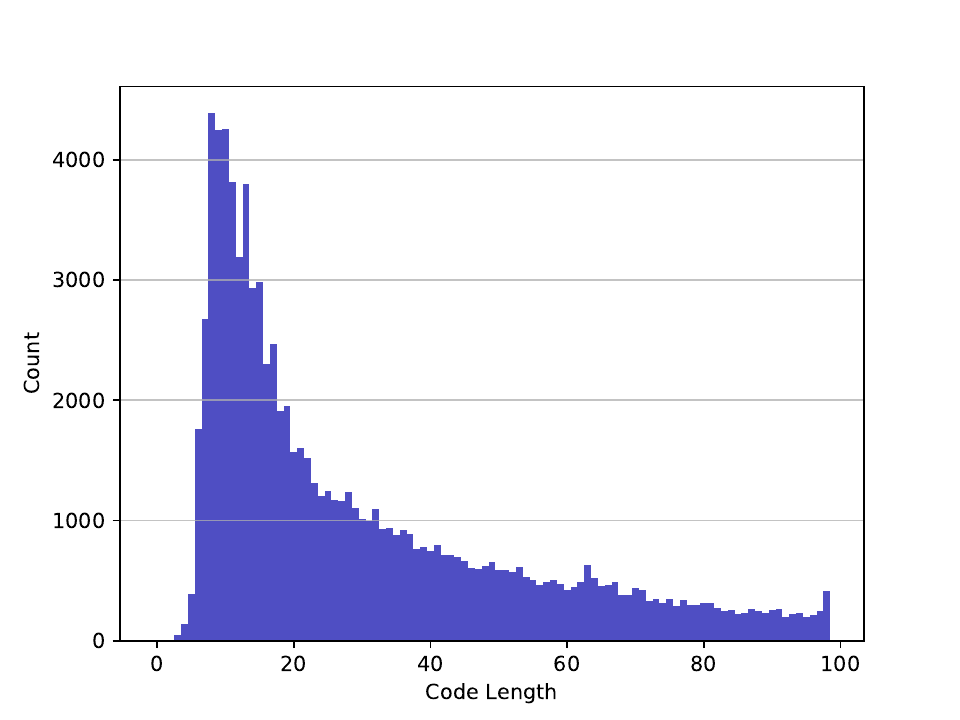}
    \caption{Code length distribution.}
    \end{subfigure}% 
    \begin{subfigure}[t]{0.5\linewidth}
    \centering
    \includegraphics[width=\textwidth]{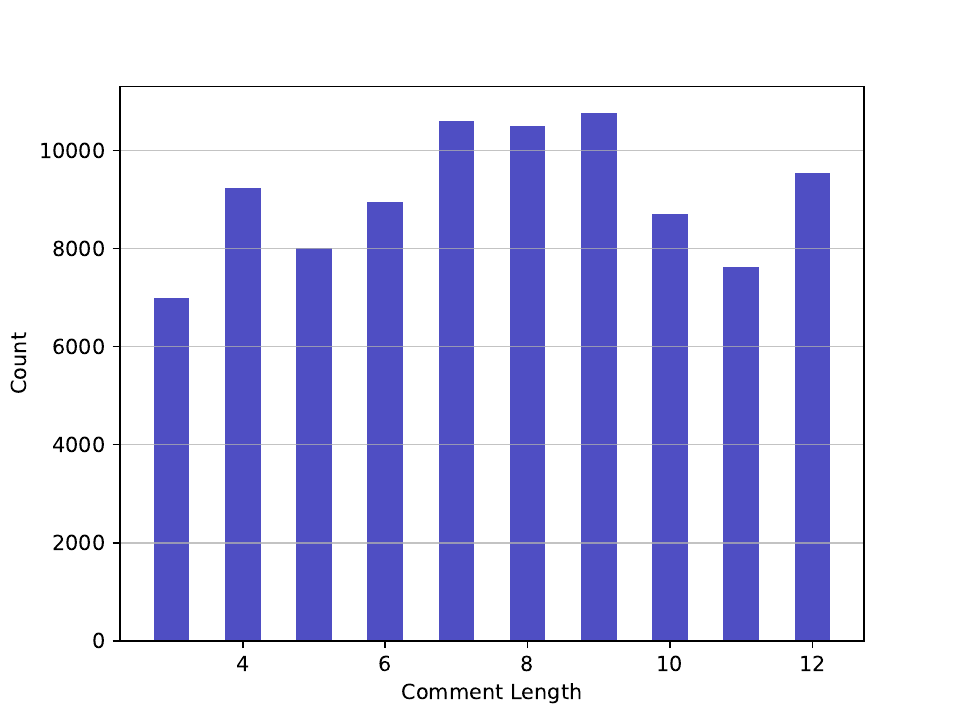}
    \caption{Summary length distribution.}
\end{subfigure}
\end{minipage}
\caption{Length distribution of test data.}
\label{fig:length_dis}
\end{figure}

\section{Experimental Setup}
\label{sec:4}
\subsection{Dataset}
Following previous studies \cite{[14],[15]}, we conduct experiments on a public large-scale Java dataset\footnote{http://leclair.tech/data/funcom/} provided by LeClair et al.~\cite{[14]}. The raw dataset contains 5.1 million Java methods, which is collected by Lopes et al. \cite{[28]} from the Sourcerer repository. Because the raw dataset contains a large number of samples (such as repeated and auto-generated code) that are not suitable for evaluating neural models, LeClair et al. cleaned and pre-processed the dataset.

Specifically, they first extracted Java methods and Javadocs from the source code repository. Assuming the first sentence of the Javadoc describes the method’s behavior \cite{[29]}, they extracted the first sentence of the Javadoc as the summary of a method and filtered out non-English samples. Considering the auto-generated and duplicate code might affect the evaluation, they removed these samples using heuristic rules \cite{[30]} and added unique, auto-generated code to the training set. After that, they split camel case and underscore tokens and set them to lower case. Finally, they divided the dataset by project into training, validation, and test set, meaning that all methods in one project are grouped into one set. They argued that the pre-processing of the dataset is necessary for evaluating the performance of neural models. The statistical results of the dataset are shown in Table \ref{tab:dataset}. Figure \ref{fig:length_dis} shows the length distribution of source code and summary on the test set. 
Based on this processed dataset, we construct new instances with the strategies described in Section~\ref{sec:3.1} for our Edit module. Finally, we can obtain 19,714,828 instances for training, 104,273 instances for validation, and 90,908 instances for testing.

\subsection{Implementation Details}
Our model is implemented based on the Pytorch\footnote{https://pytorch.org/} framework. We set word embedding and LSTM hidden states to 300 dimensions and 512 dimensions, respectively. We set the batch size to 512 and train the model using Adam \cite{[31]} with the initial learning rate of 0.001. The learning rate is decayed with a factor of 0.95 every epoch. To mitigate overfitting, we use dropout with 0.5. To prevent exploding gradient, we clip the gradients norm by 5. According to the statistics of the dataset in Table \ref{tab:dataset} and Figure \ref{fig:length_dis}, the maximum lengths of code and summary are set to 100 and 15, respectively. The vocabulary sizes of the code and summary are 50,000 and 50,000, respectively. The out-of-vocabulary tokens are replaced by UNK. We train the model for a maximum of 30 epochs and perform an early stop if the validation performance does not improve for 5 consecutive iterations. During the testing phase, we use a beam search and set the beam size to 10. We conduct all experiments on one Nvidia V100S GPU with 32 GB memory. Each experiment is run three times and the average results are reported.

\subsection{Evaluation Metrics}
\label{sec:metrics}
Following the previous studies~\cite{[14],[15],[16]}, we evaluate all approaches using the metric BLEU~\cite{[34]}, METEOR~\cite{[42]}, ROUGE-L~\cite{[43]} and ROUGE-W~\cite{[43]}. We regard a generated summary $\hat{Y}$ as a candidate and a huamn-written summary $Y$ as a reference. 

BLEU calculates the n-gram similarity between the generated sequence and reference sequence. The BLEU score ranges from 1 to 100 as a percentage value. The higher the BLEU, the closer the candidate is to the reference.
It computes the n-gram precision of a candidate sequence to the reference:
\begin{equation}
BLEU-N=BP \cdot \exp \left(\sum_{n=1}^{N} w_{n} \log p_{n}\right)
\end{equation}
where $p_n$ is the ratio of length $n$ sub-sequences in the candidate that are also in the reference. In this paper, we report the BLEU1-BLEU4 scores. $BP$ is brevity penalty.

METEOR calculates the similarity scores between a pair of sentences by:
\begin{equation}
M E T E O R=\left(1-\gamma \cdot f r a g^{\beta}\right) \cdot \frac{P \cdot R}{\alpha \cdot P+(1-\alpha) \cdot R}
\end{equation}
where $P$ and $R$ are the unigram precision and recall, $frag$ is the fragmentation fraction. $\alpha$, $\beta$ and $\gamma$ are three penalty parameters whose default values are 0.9, 3.0 and 0.5, respectively.

ROUGE-L computes F-score based on Longest Common Subsequence (LCS). Suppose the lengths of $\hat{Y}$ and $Y$ are $m$ and $n$, then:
\begin{equation}
P_{l c s}=\frac{L C S(X, Y)}{m}, R_{l c s}=\frac{L C S(X, Y)}{n}
\end{equation}
\begin{equation}
F_{l c s}=\frac{\left(1+\beta^{2}\right) P_{l c s} R_{l c s}}{R_{l c s}+\beta^{2} P_{l c s}}
\end{equation}
where $\beta=P_{lcs} / R_{lcs}$ and $F_{lcs}$ is the value of ROUGE-L. ROUGE-W~\cite{[29]} is an improved version of ROUGE-L, which is based on Weighted Longest Common Subsequence (WLCS).

\section{Experimental Results}
\label{sec:5}
To evaluate our approach, in this section, we aim to answer the following three research questions:
\begin{itemize}
    \item RQ1: How does the {\sc EditSum} perform compared to the state-of-the-art neural baselines?
    \item RQ2: How does the {\sc EditSum} perform compared to the IR-based baselines?
    \item RQ3: Does {\sc EditSum} perform better than previous approaches for tackling the keywords problem?
\end{itemize}

\begin{table*}[t]
\small
\centering
\caption{The performance of our model compared with baselines.}
\vspace{-0.1cm}
% \resizebox{\linewidth}{!}{
\begin{tabular}{l|cccccccc}
\toprule 
Approaches & Params & BLEU1 & BLEU2 & BLEU3 & BLEU4 & METEOR &ROUGE-L &ROUGE-W\\ 
\midrule
Retrieve module & - & 32.06 & 17.83 & 14.39 & 12.87 & 28.62 & 36.82 & 25.31\\
LSI & - & 31.38 & 17.05 & 13.48 & 12.07 & 27.71 & 35.09 & 24.02\\
VSM & - & 31.91 & 17.52 & 14.02 & 12.70 & 28.26 & 36.21 & 24.81\\
NNGen & - & 33.48 & 18.86 & 14.99 & 13.44 & 29.97 & 38.57 & 26.07\\
\midrule
CODE-NN & 36.3M & 32.23 & 14.71 & 8.558 & 6.090 & 29.35 & 37.64 & 25.85\\
DeepCom & 37.9M & 31.88 & 16.02 & 10.10 & 7.491 & 31.79 & 42.45 & 28.51\\
attendgru & 37.7M & 39.00 & 22.02 & 14.87 & 11.27 & 36.42 & 48.95 & 27.96\\
ast-attendgru & 39.7M & 39.32 & 22.19 & 14.98 & 11.42 & 36.99 & 49.40 & 33.58\\
Rencos & 57.3M & 34.40 & 19.82 & 14.34 & 11.74 & 34.53 & 46.35 & 31.64\\
Re$^2$Com & 28.4M & 41.69 & 25.78 & 19.70 & 16.79 & 38.04 & 47.65 & 33.22\\
{\sc EditSum} & 26.4M & \textbf{45.83} & \textbf{28.37} & \textbf{21.17} & \textbf{16.88}  & \textbf{42.93} & \textbf{53.17} & \textbf{37.19} \\
\bottomrule 
\end{tabular}
\label{tab:baselines}
\vspace{-.4cm}
\end{table*}

\subsection{RQ1: {\sc EditSum} vs. Neural Baselines}
\subsubsection{Baselines}
To answer this research question, we compare our approach {\sc EditSum} to six state-of-the-art neural models.
\begin{itemize}
    \item \textbf{CODE-NN} \cite{[12]} is the first neural network-based model for code summarization task. It maps the source code sequence into word embeddings, then uses an LSTM unit as a decoder to generate summaries, and employs the attention mechanism to introduce information from the word embeddings.
    \item \textbf{DeepCom} \cite{[13]} is a seq2seq model with an attention mechanism that uses LSTM units as the encoder and decoder. It proposed a SBT algorithm to convert the AST into a token sequence. It is the first model to introduce structural information of source code into code summarization.
    \item \textbf{attendgru} \cite{[14]} is an encoder-decoder model with an attention mechanism, which takes the code sequence as input and the summary as output.
    \item \textbf{ast-attendgru} \cite{[14]} is also a seq2seq model with an attention mechanism. Different from attendgru, it introduces the structural information of the source code by using an encoder to process the traversal sequence of AST. It concatenates the information from the two encoders as input to the decoder and generates code summaries.
    \item \textbf{Rencos} \cite{[16]} is a retrieval-based neural model that augments an attentional encoder-decoder model with the retrieved two most similar code snippets for better source code summarization.
    \item \textbf{Re$^2$Com} \cite{[15]} is a retrieval-based neural model that uses the summary of the similar code snippet as an exemplar to generate a code summary.
\end{itemize}
For a fair comparison, we re-implement the attendgru and ast-attendgru based on LSTM units. The embedding size and LSTM states of all baselines are set to 256 dimensions.

\subsubsection{Results}
We calculate the BLEU, METEOR, and ROUGE scores between the summaries generated by different approaches and human-written summaries. The experimental results are shown in Table \ref{tab:baselines}. 
We notice that CODE-NN performs the worst among all approaches. This is because CODE-NN directly uses word embeddings as the input of decoder, and does not further extract the semantic information from the source code. This shows that whether the semantic information of the code can be effectively mined has a greater impact on the performance of the code summarization model.
Both DeepCom and attendgru use the encoder-decoder framework, but DeepCom performs worse. This is because the traversal sequence of the AST input by DeepCom is about 7 times longer than the token sequence of code input by attendgru. This also verifies the weakness of LSTM in processing long sequences \cite{[35]}.
The difference between ast-attendgru and attendgru is that the former introduces the structural information in the AST, but the improvement is limited. This is because custom identifiers are removed from the AST used in ast-attendgru, which limits the structural information in the AST.
Both Rencos and Re$^2$Com combine the information retrieval technology with neural networks, but the former is less effective. Rencos retrieved two similar code snippets from the corpus and directly used them as input to the model. Re$^2$Com retrieved a similar code from the corpus, and then sent the summary of the similar code into the model as an exemplar. The experimental results show that the summary of similar code contains more valuable and reusable information than similar code that may contain noise. This also proves that it is reasonable for us to use the summaries of similar code as the prototypes.

From Table \ref{tab:baselines}, we can see that {\sc EditSum} performs the best among all neural models, which improves the state-of-the-art Re$^2$Com by 9.93\% in BLEU1, 12,85\% in METEOR and 11.58\% in ROUGE-L. 
In particular, compared with Rencos and Re$^2$Com, {\sc EditSum} performs much better on all metrics. This is because Rencos and Re$^2$Com are the ensemble neural models, and they directly enter the retrieved results and the input code into the model.
While {\sc EditSum} regards the prototype summary as an initial draft for post-generation, which provides many reusable patternized words.
So, {\sc EditSum} focuses on learning how to revise the prototype based on the differences between the input code and the similar code. Besides, the number of parameters of {\sc EditSum} is the smallest among all baselines. It also shows {\sc EditSum} is efficient. 

Compared to other metrics, we find that {\sc EditSum} has a small improvement on BLEU4. This is because the improvement by {\sc EditSum} mainly comes from predicting more keywords. However, the average length of the summaries in the test set is 7, and these keywords are mainly 1-3 words. Therefore, {\sc EditSum} has a great improvement on BLEU1-BLEU3, and a relatively small improvement on BLEU4.

\subsection{RQ2: {\sc EditSum} vs. IR Baselines}
\subsubsection{Baselines}
To answer this research question, we compare our approach {\sc EditSum} to four IR-based baselines.
\begin{itemize}
    \item \textbf{Retrieve module} is a component of our approach, whose details are described in Section \ref{sec:3.1}. We use the summary of similar code as output directly.
    \item \textbf{Latent Semantic Indexing} (LSI) \cite{[7]} is an IR technique to analyze the semantic relationship between terms in documents. Given a code snippet, we use LSI to retrieve the similar code from the training set and use its summary as output. The retrieval metric is the cosine distance of the 500-dimensional LSI vector of the source code.
    \item \textbf{Vector Space Model} (VSM) \cite{[7]} represents the code as a vector using Term Frequency-Inverse Document Frequency (TF-IDF). It uses cosine similarity to retrieve the summary of the similar code from the training set.
    \item \textbf{NNGen} \cite{[32]} is an approach for generating commit messages based on nearest neighbors algorithm. It first encodes code changes into the form of a "bag of words", then uses the cosine distance to select the nearest $k$ code changes. Finally, it chooses the message of the code change with the highest BLEU score as the final result. In this paper, we set $k$ as 5.
\end{itemize}

\subsubsection{Results}
We calculate the BLEU, METEOR, and ROUGE scores between the summaries generated by different IR-based approaches and human-written summaries. The experimental results are shown in Table \ref{tab:baselines}.
From Table \ref{tab:baselines}, the Retrieve module performs better compared with other IR-based approaches. This shows that it is effective for our Retrieve module to retrieve similar code based on the lexical similarity.
LSI and VSM use different ways (LSI vectors and TF-IDF) to map source code into a vector, and their performance is similar.
Compared with LSI and VSM, NNGen directly uses BLEU score as the retrieval metric, so it gets a higher BLEU score.
It is worth noting that the BLEU3 and BLEU4 score of the IR-based baselines even exceeds that of some neural models (i.e, CODE-NN and DeepCom). This shows that the summaries output by the IR-based approaches have better precision scores of the 3-gram phrase and 4-gram phrase.
This proves that the retrieved summaries contains a lot of valuable words, which can be used to generate the new summaries.
However, there is still a gap between the summaries output by the IR-based approaches and the human-written summaries due to the differences between the similar code and the input code.

Our model signiﬁcantly outperforms IR-based baselines in terms of all metrics, which proves the effectiveness of the our Edit module.
Compared to the IR-based baselines. our approach {\sc EditSum} treats the retrieved summary as a prototype, and then revise the prototype conditioning on the semantic differences between similar code and input code. By combining the advantages of neural networks and IR-based approaches, {\sc EditSum} achieves the best performance.

\subsection{RQ3: Tackling keywords problem}

\begin{table}[t]
\small
\centering
\caption{The number of correctly generated low-frequency words (the rate of keywords in parentheses)}
\begin{tabular}{l|cccc}
\toprule \
Approaches & $\leq$10 & $\leq$20 & $\leq$50 & $\leq$100\\ \midrule
ast-attendgru & 262 & 624 & 1,575 & 2,801  \\
Rencos   & 410  & 948  & 2,254  & 3,791  \\
Re$^2$Com & 422 (64.69\%) & 1,093 (75.21\%) & 2,808 & 4,886 \\
{\sc EditSum} & \textbf{476 (74.58\%)} & \textbf{1270 (86.38\%)} & \textbf{3066} & \textbf{5260} \\
\bottomrule 
\end{tabular}\vspace{-.2cm}
\label{tab:low-freq}
\end{table}

\subsubsection{Metrics}
According to the information retrieve technologies~\cite{[39]}, the keywords in the summaries often are informative and are more likely to be low-frequency words.
The statistics show 94.8\% of tokens in the summary vocabulary of the dataset have a frequency of less than 100. However, as we described in Section \ref{sec:1} and \ref{sec:2}, previous neural network models perform poorly on predicting low-frequency words. 
To measure the ability of our approach on generating keywords, we collect all correctly predicted words on the test set, calculate the frequency of these words on the training set, and count the words with frequencies less than 10, 20, 50, and 100. The correctly predicted words refers to the overlap between the generated summary and the reference summary. For the words with frequencies less than 10 and 20, we manually counted the rate of keywords among these words.

\subsubsection{Results}
The statistical results are shown in Table \ref{tab:low-freq}.
Compared with ast-attendgru, Rencos and Re$^2$Com perform better on predicting the low-frequency words. This shows that the summaries of similar code snippets contain a lot of reusable information. 
We also can see that {\sc EditSum} can predict more low-frequency words and more keywords than other baselines, which means that {\sc EditSum} alleviates the problem of predicting keywords. 
The goal of learning how to revise a prototype makes our model focuses to generate more keywords.

\subsection{Human Evaluation}

\begin{table}[t]
\small
\centering
\caption{The results (standard deviation in parentheses) of human evaluation.}
\vspace{-0.2cm}
\resizebox{\linewidth}{!}{
\begin{tabular}{l|ccc}
\toprule \
Approaches & Naturalness & Informativeness & Usefulness\\ \midrule
Retrieve module & 1.790 (0.68) & 0.778 (0.59) & 0.612 (0.12) \\
ast-attendgru & 1.713 (0.76) & 1.288 (0.79) & 1.108 (0.89) \\
Rencos   & 1.822 (0.73) & 1.320 (0.36) &  1.140 (0.29) \\
Re$^2$Com & 1.860 (0.64) & 1.465 (0.52) & 1.341 (0.23) \\
{\sc EditSum} & \textbf{1.933} (0.31)  & \textbf{1.802} (0.348)  & \textbf{1.790} (0.29) \\
\bottomrule 
\end{tabular}}
\label{tab:human}
\end{table}

\subsubsection{Metrics}
Although the metrics in Section \ref{sec:metrics} can calculate the lexical similarity between the generated summaries and the reference summaries, they can not reflect the similarity at the semantic level. Moreover, the ultimate goal of the automatic code summarization model is to help developers understand the functionality of the program. Therefore, we conduct a human evaluation to measure the quality of summaries generated by different baselines on the test set. Following the previous work \cite{[15]},
we measure three aspects, including the \textit{naturalness} (grammaticality and fluency of the generated summaries), \textit{informativeness} (the amount of content carried over from the input code to the generated summaries, ignoring fluency of the text), and \textit{usefulness} (what extent the generated summary is useful for developers to understand code). All three scores are integers, ranging from 0 to 2 (from bad to good). We invite 10 volunteers with 3-5 years of Java development experience and excellent English ability for 1 hour each to evaluate the generated summaries in the form of a questionnaire. The 10 volunteers are computer science Ph.D. students and are not co-authors of this paper. We randomly select 500 samples generated by five models (100 from Retrieve module, 100 from ast-attendgru, 100 from Re$^2$Com, 100 from Rencos, and 100 from {\sc EditSum}). The 500 samples are divided into five groups, with each questionnaire containing one group. We randomly list the summary pairs and the corresponding input code on the questionnaire and remove their labels. Each group is evaluated by two volunteers, and the final result of a pair of summaries is the average of two volunteers. Volunteers are allowed to search the Internet for related information and unfamiliar concepts.

\subsubsection{Results}
The evaluation results are shown in Table \ref{tab:human}. The standard deviations of all approaches are small, indicating that their scores are about the same degree of concentration. Our model is better than all baselines in three aspects. The Retrieve module can generate more fluent summaries than the ast-attendgru because its outputs are directly retrieved from the training set. We also notice that the scores on \textit{informativeness} of five models are higher than those on \textit{usefulness}. This indicates that the generated 
summaries really contain information about the code, but this information may be redundant or not completely correct, so they only provide limited help for developers to understand the code.

\section{Discussion}
\label{sec:6}
\subsection{Qualitative Analysis}
\begin{table}[t]
\centering
\scriptsize
\caption{Examples of generated summaries.}
\vspace{-.2cm}\begin{tabular}{c|l}
\toprule
Case ID & Example  \\
\midrule
\multirow{7}{*}{1} &
\begin{lstlisting}
public void drawSelected(){
    if(unselectedColor instanceof Color){
      setPaint(((Color)unselectedColor).darker());
    }else{
      setPaint(java.awt.Color.yellow);
    }
}
\end{lstlisting}\\
&\textbf{Retrieve Module}: set the series colors to the chart \\ 
&\textbf{ast-attendgru}: draws the selected set of the selected color \\
&\textbf{Rencos}: p method description p \\
&\textbf{Re$^2$Com}: set the selected color to the specified color \\
&\textbf{{\sc EditSum}}: set the color to a darker shade \\
&\textbf{Human-written}: set the color to a darker shade \\
\midrule
\multirow{7}{*}{2} &
\begin{lstlisting}
public void close() throws IOException {
    this.servletInputStream.close();
}
\end{lstlisting}\\
&\textbf{Retrieve Module}: close the resources used by the work factory \\ 
&\textbf{ast-attendgru}: close the underlying servlet \\
&\textbf{Rencos}: close the server \\
&\textbf{Re$^2$Com}: close the resources used by the work factory \\
&\textbf{{\sc EditSum}}: close the underlying stream \\
&\textbf{Human-written}: close the underlying stream \\ 
\midrule
\multirow{7}{*}{3} &
\begin{lstlisting}
public int read() throws IOException{
    if(chunkSize==-1){
        return -1;
    }
    if(pos<chunkSize){
        pos++;
        return in.read();
    }else{
        readChunksize();
        pos=0;
        if(chunkSize<=0){
            return -1;
        }
        pos=1;
        return in.read();
    }
}
\end{lstlisting}\\
&\textbf{Retrieve Module}: read some bytes from the stream \\ 
&\textbf{ast-attendgru}: reads the next byte \\
&\textbf{Rencos}: reads the next byte \\
&\textbf{Re$^2$Com}: read some bytes from the stream \\
&\textbf{{\sc EditSum}}: read the next byte of data from this input stream \\
&\textbf{Human-written}: read the next byte of data from this input stream \\
\bottomrule
\end{tabular}\vspace{-.4cm}
\label{tab:case}
\end{table}

We present three examples generated by different approaches from the test set, as shown in Table \ref{tab:case}. 
These examples show that the summaries generated by {\sc EditSum} have a very high semantic similarity with human-written summaries.
From Table \ref{tab:case}, previous models cannot generate keywords accurately, and the generated summaries cannot reflect the intention of the programs concisely. 
For example, in case 1, the aim of this program is to set the color to a darker shade. However, Re$^2$Com simply describes it as setting the selected color to the specified color, which is useless for developers to understand the program. 
While our model {\sc EditSum} performs well in both patternized words (e.g. set, to) and keywords (e.g. darker shade).
Besides, compared with Retrieve module, we can find that our Edit module can make good use of the pattern in the prototype and revise it based on the semantics of the input code.

\subsection{Performance for Different Lengths}

\begin{figure}[t]
\centering
\includegraphics[width=\linewidth]{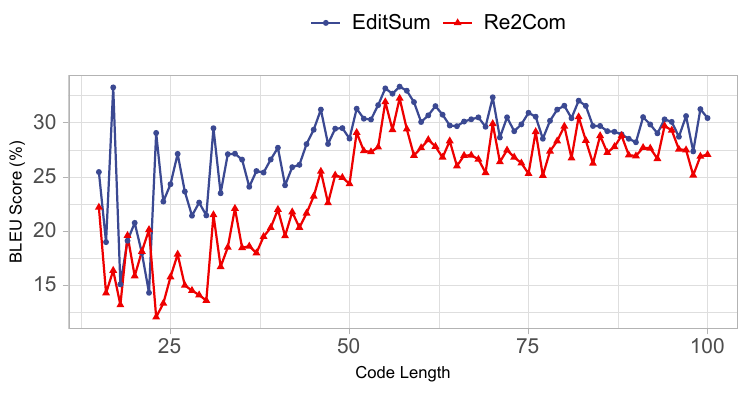}
\caption{BLEU scores for different code lengths.}
\label{fig:code_length_bleu}
\end{figure}

\begin{figure}[t]
\centering
\includegraphics[width=\linewidth]{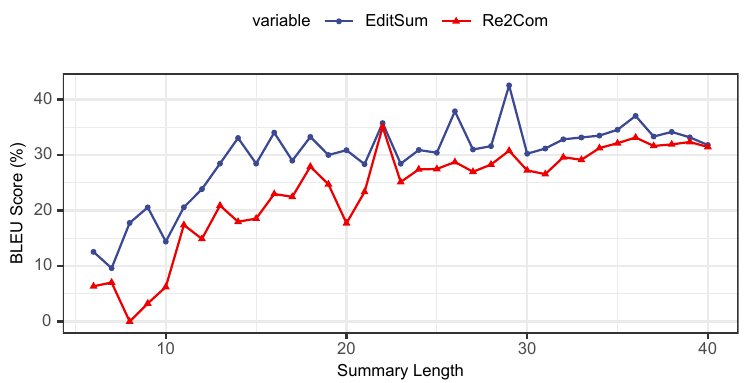}
\caption{BLEU scores for different summary lengths.}
\label{fig:summ_length_bleu}
\end{figure}

We also analyze the performance of different models on different code and summary lengths (number of tokens). We calculate the BLEU score for each sample on the test set and average the scores by length. The experimental results are shown in Figure \ref{fig:code_length_bleu} and Figure \ref{fig:summ_length_bleu}. 
From these figures, we can observe that {\sc EditSum} outperforms the Re$^2$Com with different code and summary lengths.
As the lengths of the code and summary increase, {\sc EditSum} keeps a stable improvement over Re$^2$Com. 
The performance of our model is always better than the baseline on the complicated code snippets with a relatively large length.
This also shows the robustness of our model.

\subsection{Threats to Validity}
There are three main threats to the validity of our model.
Firstly, we only conducted experiments on a Java dataset. Although Java may not be representative of all programming languages, the experimental dataset is large and safe enough to show the effectiveness of our model. Previous studies \cite{[15],[16]} also only conducted experiments on this Java dataset.
Besides, our model uses only language-agnostic features and can be applied in a drop-in fashion to other programming languages. 
Secondly, we cannot guarantee that the scores of human evaluation are fair. To mitigate this threat, we evaluate every code-summary pair by two evaluators and use the average score of the two evaluators as the final result. 
Finally, the Retrieve module retrieves similar code based on lexical similarity. This may result in retrieved code and input code being similar only at the lexical level, but their summaries are quite different.
To address this threat, we use a large-scale Java dataset (2M) to increase the scale and diversity of retrieval corpus. 
The experimental results in Table \ref{tab:baselines} prove that the performance of our retrieval module is comparable to the performance of some neural network models (CODE-NN, DeepCom).
We also propose an Edit module to alleviate this threat through revising the prototype according to the semantic differences between input code and retrieved code.

\section{Related Work}
\label{sec:7}
As an integral part of software development, code summaries describe the functionalities of source code. A concise and clear summary can help developers quickly understand the purpose of the program. However, it is very time-consuming and labor-intensive to write a summary manually.
Therefore, more and more researchers are exploring automatic code summarization technology.
Automatic code summarization approaches vary from manually-crafted templates \cite{[4],[5],[17],oda2015learning}, information retrieval \cite{[6],[7],[8],[9]} and neural networks \cite{[12],[13],[14],[15],[16]}.

\subsection{Template-based Approaches}
Early studies generated code summaries based on template-based approaches.
Given the signature and body of a method, Sridhara et al. \cite{[5]} identified the content for the summary and generated natural language text that summarizes the method's actions.
Moreno et al. \cite{[4]} determined the class and method stereotypes and used them, in conjunction with heuristics, to select the information to be included in the summaries. Then they generated the summaries using existing lexicalization tools.
McBurney et al. \cite{[17]} utilized the PageRank algorithm~\cite{pagerank:2004} to select the important methods in the given method’s context and used a template-based system to generate English descriptions of Java methods.
Generating summaries based on templates can improve the readability of summaries, but defining templates is a time-consuming task and requires extensive domain knowledge. Besides, templates of different projects cannot be migrated to each other.

\subsection{IR-based Approaches}
Information retrieval technologies are also widely used in automatic code summarization. 
Haiduc et al. \cite{[7]} represented the source code as a vector using two algorithms (VSM and LSI) and retrieved relevant terms from a code corpus. These relevant terms were integrated into a code summary.
Eddy et al. \cite{[6]} proposed a hierarchical probabilistic model that retrieved relevant terms from the code corpus and fused them into the summaries.
Wong et al. \cite{[9]} applied a token-based code clone detection tool to retrieve similar code snippets in large-scale software repositories.
Although promising, IR-based approaches have two main limitations: first, they fail to extract accurate keywords used to identify similar code snippets when identifiers and methods are poorly named. Second, they rely on the size and diversity of the retrieval corpus.

\subsection{Neural Network-based Approaches}
Recently, more and more neural networks are applied to generate code summaries. 
Iyer et al. \cite{[12]} used a Recurrent Neural Network (RNN)~\cite{[48]} with an attention mechanism as a decoder to generate code summaries and achieved good results on C\# and SQL summaries.
Because source code contains rich structural information, Hu et al. \cite{[13]} proposed a neural model named DeepCom to utilize the structural information of code. They proposed a SBT algorithm to convert the AST into a token sequence, then designed a seq2seq model to generate summaries for Java methods based on the AST sequence.
LeClair et al. \cite{[14]} proposed two neural models (attendgru and ast-attendgru) to generate the summaries by combining the sequence information and structure information of the code. 
Wei et al. \cite{[15]} proposed an exemplar-based summary generation framework that used the summary of the similar code snippet as an exemplar to assist in generating a target summary.
Zhang et al. \cite{[16]} proposed a retrieval-based neural model that augments an attentional seq2seq model with the retrieved two most similar code snippets for better source code summarization.

Different from the retrieval-based neural models \cite{[15],[16]}, we regard the retrieved summary as a prototype and combine the pattern in prototype with semantic information of input code.
While previous models formulate it as a multi-source seq2seq task, in which the input code, prototype, and similar code are all fed to the decoder. The experimental results also prove the superiority of our approach.

\section{Conclusion and Future Work}
\label{sec:8}
In this paper, we argue that code sumaries are composed of patternized words and keywords, and emphasize the shortcomings of previous models in predicting keywords.
To alleviate this problem, we propose a retrieve-and-edit approach named {\sc EditSum} for code summarization. {\sc EditSum} contains two modules. A Retrieve module for retrieving the similar code snippet. An Edit module treats the summary of similar code as a prototype, and combine the pattern in prototype and semantic information of input code to generate a target summary.
We conducted extensive experiments on a large-scale Java dataset. The experimental results show that {\sc EditSum} substantially outperforms the state-of-the-art neural baselines and the IR-based baselines. Human evaluation and case analysis prove that {\sc EditSum} can generate concise and informative summaries, which can effectively help developers understand the intent of the programs. 
The analysis of the experimental results shows that {\sc EditSum} can generate more keywords and performs well on code and summaries of different lengths. 
In the future, we will explore how to generate standard and meaningful code summaries for various software projects.

\section*{ACKNOWLEDGMENTS}
This research is supported by the National Key R\&D Program of China under Grant No. 2020AAA0109400, and the National Natural Science
Foundation of China under Grant Nos. 62072007, 61832009, 61620106007.

\normalem
\bibliographystyle{IEEEtran}
\bibliography{IEEEabrv,mybibfile}

\end{document}